\documentclass[sigconf]{acmart}
\usepackage{makecell}
\usepackage{graphicx}
\usepackage{booktabs}
\usepackage{pifont}% http://ctan.org/pkg/pifont
\usepackage{tcolorbox}
\usepackage{caption}
\usepackage{subcaption}
\usepackage{tcolorbox}
\usepackage{array}
\usepackage[ruled, lined, linesnumbered, commentsnumbered, longend]{algorithm2e}
\usepackage{xcolor}
\usepackage{url}

\newcommand{\cmark}{\ding{51}}%
\newcommand{\xmark}{\ding{55}}%

\newcommand{\newtext}[1]{#1}

\def \ori {$D_{ori}$\xspace}
\def \neg {$D_{rep}$\xspace}
\def \aug {$D_{aug}$\xspace}
\def \bla {$D_{bl1}$\xspace}
\def \blb {$D_{bl2}$\xspace}
\def \blc {$D_{bl3}$\xspace}
\def \bld {$D_{bl4}$\xspace}
\def \blx {$D_{bl*}$\xspace}

\title{Too Few Bug Reports? Exploring Data Augmentation for Improved Changeset-based Bug Localization}

\author{Agnieszka Ciborowska}
\affiliation{%
  \institution{Virginia Commonwealth University}
  \city{Richmond, Virginia}
  \country{USA}
}
\email{imranm3@vcu.edu}

\author{Kostadin Damevski}
\affiliation{%
  \institution{Virginia Commonwealth University}
  \city{Richmond, Virginia}
  \country{USA}
}
\email{kdamevski@vcu.edu}

\begin{abstract}
Modern Deep Learning (DL) architectures based on transformers (e.g., BERT, RoBERTa) are exhibiting performance improvements across a number of natural language tasks. 
While such DL models have shown tremendous potential for use in software engineering applications, they are often hampered by insufficient training data. Particularly constrained are applications that require project-specific data, such as bug localization, which aims at recommending code to fix a newly submitted bug report. 
Deep learning models for bug localization require a substantial training set of fixed bug reports, which are at a limited quantity even in popular and actively developed software projects. 
In this paper, we examine the effect of using synthetic training data
on transformer-based DL models that perform a more complex variant of bug localization, which has the goal of retrieving bug-inducing changesets for each bug report. To generate high-quality synthetic data, we propose novel data augmentation operators that act on different constituent components of bug reports. We also describe a data balancing strategy that aims to create a corpus of augmented bug reports that better reflects the entire source code base, because existing bug reports used as training data usually reference a small part of the code base.
Data balancing helps the model perform better for newly-reported bug reports that reference previously unobserved code. Our evaluation results indicate that both data augmentation and balancing are effective, improving retrieval performance across all three BERT-based models we studied.
\end{abstract}

\begin{document}

\maketitle

\section{Introduction}\label{sec:introduction}

% DL in SE
The emergence of novel Deep Learning (DL) architectures, such as transformers, has fueled outstanding improvements across multiple tasks in Natural Language Processing (NLP), and encouraged their application to various problems in the software engineering domain. Software engineering researchers have studied the potential of DL in the context of problems such as code search~\cite{husain2019codesearchnet,lin2021traceability,gu2018deep,guo2017semantically}, defect prediction~\cite{7272910, li2017software,hoang2019deepjit,wang2018deep, pornprasit2022deepline}, and bug localization~\cite{deeptransfer,ciborowska2021fast,lam_dnnloc,huo2016learning,xiao2019improving}.
However, the fundamental weakness of DL approaches is that they require large amount of labelled data to train the model. 
At the same time, maintaining the quality of the labelled data is crucial to achieve the best performance.
While manual labelling is typically a preferred approach to ensure high data quality, it is a slow and time-consuming process~\cite{tu2020better}, often intractable considering the amount of data required to train a DL model. On the other hand, automated mining for labels is far more likely to meet the demand for data quantity, however at the cost of introducing noise in the form of both false positives and false negatives~\cite{vasilescu2015qualilty,dacosta2017framework}. 
Hence, collecting large amount of good quality labelled data can pose a significant challenge for many important software engineering problems and tasks, in particular those that require single project data (i.e., within project)~\cite{tu2021frugal}. 
A recent approach to address this problem is to use transfer learning, i.e., pre-training a model with unsupervised learning on a large, general corpus, followed by fine-tuning via supervised learning towards the downstream task. 
However, this strategy still requires a non-trivial dataset for fine-tuning and, as observed by Gururangan et al., it leads to suboptimal performance compared to when a model is pre-trained and fine tuned on in-domain data~\cite{gururangan2020don}.

% However, the fundamental weakness of DL approaches is that they require large amount of labelled data to train the model. At the same time, maintaining the quality of the labelled data is crucial to achieve the best performance. 
% Collecting good quality labelled data in the required quantity can pose a significant challenge for many important software engineering problems and tasks, in particular those that require single project data (i.e., within project)~\cite{tu2021frugal}. 
% Manual labelling is a slow and time-consuming process, often intractable considering the amount of data required to train a DL model. Automated mining for labels can introduce noise in form of both false positives and false negatives. 
% A recent approach to address this problem is to use transfer learning, pre-training a model with unsupervised learning on a large, general corpus, followed by fine-tuning via supervised learning towards the downstream task. However, as observed by Gururangan et al. this strategy leads to suboptimal performance compared to when a model is trained on too small in-domain data~\cite{gururangan2020don}. 
%Similarly, Lin et al.~\cite{lin2021traceability} noted that pre-training a model on data from closely related domain improves the results.

% Bug Localization
% TODO: explain that we do bug-inducing changeset BL.
One of the software engineering tasks that benefits from a DL-based approach is bug localization, which aims to identify relevant code entities (e.g., classes, methods or changesets) for a given bug report describing a software failure. Over the years, researchers have proposed multiple approaches for bug localization based on the Vector Space Model (VSM)~\cite{saha_bluir_2013,wang_amalgam_2014,wen2016} and probabilistic models (e.g., Latent Dirichlet Allocation)~\cite{corley2018,bugscout}, while also recognizing that the key drawback of those techniques is their limited ability to deal with the semantic gap between source code causing the bug and the description given in the bug report~\cite{bettenburg2008what,ye2014learning}. To address that, recent efforts have been diverted towards DL techniques, including RNN, LSTM and, finally, transformer-based models~\cite{lin2021traceability,ciborowska2021fast}. As noted by Guo et al.~\cite{guo2017semantically}, the availability of training data is one of the key factors limiting DL performance. In the case of bug localization, the training data consists of pairs of bug reports and their introducing (or inducing) changesets, which are difficult to obtain at scale for a couple of key reasons. First, matching a bug report to bug-introducing changesets is challenging as developers rarely mark culprit code changes explicitly~\cite{murali2020industry}, while approaches that find the bug-inducing changesets automatically, based on the SZZ algorithm~\cite{sliwerski2005when}, are prone to introducing noise~\cite{rosa2021evaluating,neto2018impact}. 
Second, the number of positive samples is bounded by the number of fixed bug reports, which are limited even for large and actively maintained projects. Relatively smaller software projects with, e.g., dozens of fixed bug reports, would be very difficult to use.
In the end, the main question remains open: \newtext{how to leverage DL techniques for bug localization, given the paucity of project-specific data.}

% DA
% solving imbalance problem != doing DA
In the NLP domain, this question has been answered with some success by Data Augmentation (DA) techniques, which, in general, can be described as strategies to artificially increase the number and diversity of training samples based on the currently available data~\cite{feng2021survey}. DA aims to create high quality synthetic data by applying transformations to the available data, while maintaining label invariance. As a result, the size of the original dataset increases, which in turn enables training a DL model for low resource domains and tasks.

% This work
%Encouraged by recent advances of DA in the NLP domain, in this work, we aim to explore data augmentation in the context of DL-based models for bug localization. 
Encouraged by recent advances of DA in the NLP domain, in this work, we aim to explore data augmentation for bug reports with the goal of producing a large number of high quality, realistic, synthetic bug reports, which can be subsequently used to increase the size of the training set for a bug localization DL model.
To this end, we propose two sets of DA operators that independently target natural language text and code-related data (e.g., code tokens, stack traces and code snippets) in each bug report.  
More specifically, natural language text is augmented using token- and paragraph-level transformations (e.g., synonym inserts), while the code-related data is augmented with code tokens from its respective bug-inducing changesets in order to strengthen the connection between a bug report and different portions of its introducing changeset.
%%% realistic looking data, diversity and quality
%More specifically, we generate significantly more training data by augmenting existing bug reports using a set of DA operators.
At the same time, by leveraging the augmented bug reports we plan to achieve another important goal, balancing the augmented dataset toward parts of the source code underrepresented in the original training set. \newtext{This addresses the common occurrence in software projects that existing bug reports reference only a specific part of the code base, while other parts have few or no bug reports, leading to the bug localization model overly focusing only on a part of the code base.}
In this paper, we investigate the following Research Questions (RQs):

\medskip

\noindent
\textbf{RQ1: }{\em (a) Can Data Augmentation improve the retrieval performance of DL-based bug localization? (b) How does Data Augmentation impact the performance of different DL-based bug localization approaches?}

\smallskip
\noindent
To understand whether Data Augmentation is a relevant strategy in DL-based bug localization, we identify three recent transformer-based models to perform this task. We evaluate the performance of these bug localization approaches, with and without DA, using a standard bug localization dataset and metrics commonly used to measure information retrieval performance. As augmentation necessarily introduces significantly higher data quantity, we add baselines to the evaluation that aim to differentiate the quantity vs. the quality of the augmented dataset. The results indicate that (1) the proposed data augmentation strategy improves retrieval accuracy \newtext{by between 39\% and 82\%},
%9.5 to 14.8 percentage points; 
(2) augmenting the dataset is more beneficial than increasing the size of the training dataset by repetition; and 
(3) balancing the training dataset results in improvement in retrieval performance, but the magnitude of the improvement depends on the architecture of the DL model.

\medskip

\noindent
\textbf{RQ2: }{\em Which of the proposed DA operators contribute the most to retrieval performance?} 

\smallskip
\noindent
The Data Augmentation approach in RQ1 relies on augmentation operators that perform specific types of transformations (e.g., insert, remove). In RQ2, we aim to understand the impact of these augmentation operators on the retrieval performance during bug localization. To answer RQ2, we perform ablation studies, training each DL model with augmented datasets created using all but one augmentation operator type. The results indicate that most of the operators contribute to the final performance, while certain operators are more consistent across different DL models.

% \noindent
% \textbf{RQ3:}{\em What are other scenarios DA can help?}

% \noindent
% \textbf{RQ4:}{\em How far can we extrapolate from the available data? What are the potential pitfalls of DA in the context of low resource projects?}

% The key contributions and findings of this paper can be summarized as follows:
% \begin{itemize}
%     \item DA operators for code
% \end{itemize}

% \input{background}

\section{Data augmentation for bug localization}
% Explain why we do only DA on bug reports and then follow with goal and their descriptions
With the increasing complexity of DL-based methods for bug localization~\cite{ciborowska2021fast, deeptransfer, lam_dnnloc, gupta2019neural}, the problem of data scarcity comes to the forefront. More specifically, while the more advanced models have the potential to bridge the lexical gap between a bug report and source code~\cite{bettenburg2008what, ye2014learning}, in order to fulfill that promise, they require large amount of bug reports to learn the semantics of the project and subsequently associate it with bug-inducing changesets.
Insufficient amount of training examples may lead to model overfitting,  memorizing high-frequency patterns or structures instead of generalizing the knowledge~\cite{shorten2021text}. DA can help to address the data scarcity problem in bug localization by focusing on the following goals.

%%% NEW VERSION
\noindent
\textbf{1. Increasing the number of bug reports.} 
Training a DL model for bug localization requires a substantial dataset consisting of bug reports and bug-inducing changesets. The main challenge of constructing such dataset is that it is project-specific. 
Most software projects typically have few bug reports with a clear indication of the changesets that caused them~\cite{murali2020industry}. Moreover, the total number of bug reports in a project is an upper bound on the number of positive training instances that are available. Note that while we can create numerous negative instances (i.e., a bug report and a non-bug-inducing changeset), the benefit to the DL model is limited as the bug report remains the same in each instance.
Moreover, out of all the reported bugs, some are closed with {\em Won't fix} or {\em Not a Bug} status~\cite{kochhar2014,widyasari2022influence}, hence they do not have corresponding changesets and cannot be used for training. 
To empirically verify the scale of data scarcity problem in bug localization data, we examined Bench4BL~\cite{lee2018}, a large bug localization dataset. Bench4BL includes 10K bug reports and their fixes coming from 51 popular and actively developed open source software projects, which equals to roughly 200 bug reports per project. Considering that the projects in the Bench4BL dataset are typically large and well-established (e.g., long running Apache Software Foundation projects like Camel and Hive), 200 bug reports is a discouragingly low number when it comes to ability to train an effective DL model.

\noindent
\textbf{2. Maintaining label invariance of bug reports.} 
In NLP, data augmentation is primarily evaluated on classification tasks, such as sentiment analysis or topic classification, in which rarely a single word can be representative of the overall result (i.e., a sentiment or a topic). 
Data in software engineering is a mix of natural language and code-related segments.
In case of bug localization, this mix typically affects bug reports, which often contain not only natural language description but also mentions of relevant program elements, stack traces or code snippets~\cite{bettenburg2008what}.
Applying off-the-shelf data augmentation transformations to bug localization data may cause more harm than good as it does not differentiate between NL and code, which both bring useful information, but in different forms and quantities.
Table~\ref{tab:da-example} shows examples of textual augmentation performed on the summary of bug report \#55996 from the \texttt{Tomcat} project using two augmentation operators proposed by Wei et al.~\cite{wei2019eda}. Random Swap exchanges two randomly selected words, while Synonym Replacement substitutes a randomly selected word with its synonym. To find synonyms, we use BERTOverflow~\cite{tabassum2020code}, a BERT model pre-trained on the StackOverflow corpus. Given the randomness of data augmentation operations, we see different versions of augmented bug report summary. While {\em Random Swap 1} swaps two words without affecting the semantics, {\em Random Swap 2} exchanges words that can indicate the relevant code component, if a project contains {\em AsyncContext} and {\em AsyncConnector} classes. Similarly, in the case of {\em Synonym Replacement 1} changing {\em context} to {\em session} affects the semantics less than replacing {\em Async} with {\em TCP} which are different concepts. This toy-example shows how easily off-the-shelf data augmentation can introduce noise that affects the original label, especially when handling data that contains key software engineering-related phrases. Hence augmentation of software engineering data in general, and bug reports in particular, requires additional steps to ensure the invariance of the newly generated data points.

\begin{table}[t]
\caption{Four examples of textual data augmentation with varying validity~\cite{wei2019eda}.}
\label{tab:da-example}
\begin{scriptsize}
\begin{tabular}{l|p{4.0cm}|c}
\toprule
                               & \multicolumn{1}{c|}{\textbf{Bug report summary}}                       & \multicolumn{1}{c}{\textbf{Valid}} \\ \midrule \vspace{0.1cm}
\textbf{Original}              & Async connector does not timeout with HTTP NIO context. &     --                                  \\ \vspace{0.1cm}
\textbf{Random Swap 1}         & Async connector does \underline{\textbf{timeout}} \underline{\textbf{not}} with HTTP NIO context. &  \cmark                                     \\ \vspace{0.1cm}
\textbf{Random Swap 2}         & Async \underline{\textbf{context}} does not timeout with HTTP NIO \underline{\textbf{connector}}. &  \xmark                                     \\ \vspace{0.1cm}
\textbf{Synonym Replacement 1} & Async connector does not timeout with HTTP NIO \underline{\textbf{session}}. & \cmark                                      \\
\textbf{Synonym Replacement 2} & \underline{\textbf{TCP}} connector does not timeout with HTTP NIO context. &  \xmark \\ \bottomrule                                    
\end{tabular}
\end{scriptsize}
\end{table}

\noindent
\textbf{3. Diversifying the training data.}
The goal of data diversification in DA is to ensure that augmented data introduces ''new quality'' to a training set, such as previously unobserved motifs, patterns or expressions, leading a DL model to learn the meaning behind the data instead of memorizing certain forms~\cite{nguyen2021data, shorten2021text}.
In the case of bug localization, the training dataset depicts how natural language describing a bug connects to source code concepts in the bug-inducing changeset. 
Commonly, the natural language in bug reports consists of Observed Behavior (OB), Expected Behavior (EB), or Steps to Reproduce (S2R)~\cite{chaparro_observed_2017}. Given that OB, EB and S2R have been recognized by developers as useful information when fixing a bug~\cite{bettenburg2008what}, augmentation for bug localization data should focus on introducing diversity into those through, e.g., paraphrasing their sentences. The second important component of diversification of a bug localization training set are the connections between bug reports and source code. While it is true that bugs are not evenly distributed in the source code base, the over-representation of one source code component (e.g., class, package) in the training set, may lead to the model blaming that particular component for every bug. 
To account for that, while augmenting training set for bug localization, additional steps can be taken to mitigate that risk, through, e.g., creating more augmented bug reports for those source code components that occur less often in the training set.
In summary, the diversification of training data should focus on: (1) modifying the natural language content of a bug report, and (2) diversifying how a bug report connects to the source code.

\begin{figure*}[t]
    \centering
    \includegraphics[width=0.85\textwidth]{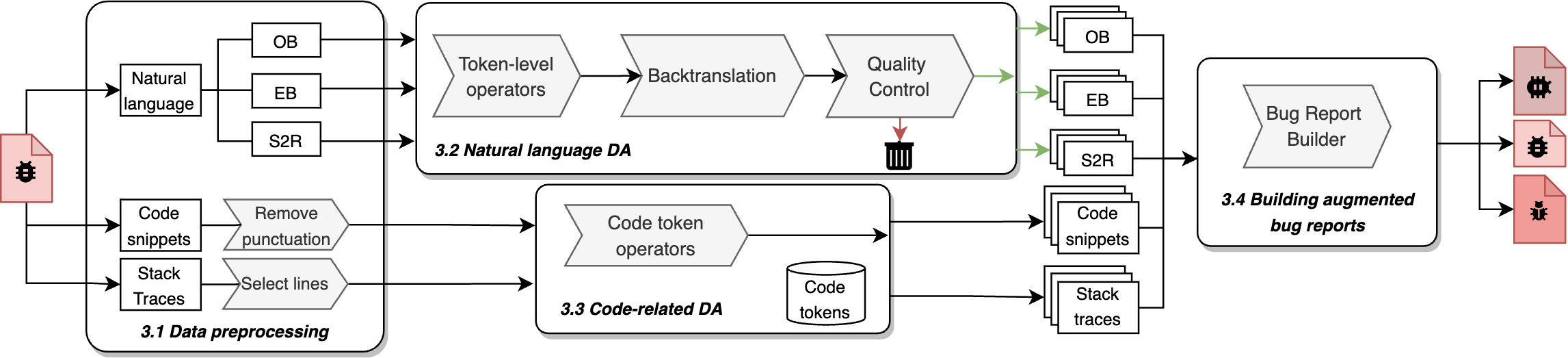}
    \caption{A visualization of our augmentation pipeline for a single bug report.}
    \label{fig:da-pipeline}
\end{figure*}

\section{Approach}

To create augmented bug reports that introduce diversity and preserve invariance (i.e., the augmented bug report still matches the same changeset as the original bug report), we propose a set of custom DA operators. %inspired by recent work in the NLP community. 
Bug reports describe software failure using various types of information, such as natural language, code snippets or stack traces, which may have different impact on matching a bug report to its inducing changeset, hence we decided to separately augment natural language and code-related information to ensure invariance of the newly created data points and avoid introducing noise. 
Figure~\ref{fig:da-pipeline} illustrates the workflow of our data augmentation process which starts with data extraction and preprocessing, followed by augmentation with the proposed operators, and construction of augmented bug reports combining the newly generated data.

% \begin{table}[t]
% \begin{scriptsize}
% \centering
% \caption{Bug report data augmentation operators.}
% \label{tab:operators}
% \begin{tabular}{l|p{3.5cm}|l}
% \toprule
% \textbf{Transformation} & \textbf{Natural Language Operator} & \textbf{Code-related Operator} \\ \midrule
% \multicolumn{3}{c}{\textit{Token-level operators}} \\ \midrule
% Insert & Dictionary Insert & Code Token Insert \\
% Replace & Dictionary Replace & Code Token Replace \\
% Swap & Random Swap & Code Token Swap \\
% Delete & Random Delete & -- \\ \midrule
% \multicolumn{3}{c}{\textit{Paragraph-level operators}} \\ \midrule
% %Backtranslation & English $\rightarrow$  German $\rightarrow$ English & -- 
% Rephrase & Backtranslation & -- 
% \\ \bottomrule
% \end{tabular}
% \end{scriptsize}
% \end{table}

\subsection{Data preprocessing}
As a first step, we use \textit{infozilla}~\cite{premraj2008extracting}, a tool that extracts stack traces and code snippets from unstructured bug report content, leaving the remaining text broadly categorized as natural language. The \textit{infozilla} produces minimal error as experiments have shown it to have 97\%+ precision, 95\%+ recall and 97\%+ accuracy. To bring out further structure from the natural language data, we extract Observed Behavior (OB), Expected Behavior (EB), and Steps to Reproduce (S2R) using the BEE tool~\cite{song2020bee}, which has shown to be highly effective at this task (94\%+ accuracy, 87\%+ recall, 70\%+ precision).

Stack traces are a valuable source of localization hints, however, due to their length they tend to introduce noise through multiple mentions of classes not necessarily related to a particular bug report~\cite{ciborowska2021fast,rahman2018improving}.
%Moreover, stack traces are typically lengthy, hence they may exceed the input size limit of commonly used DL models (e.g., BERT has an input limit of 512 tokens), while models with no input size limit will require longer computational time to process data that is likely to be just noise. 
To mitigate the noise in stack traces, we reduce their size by selecting the lines that are most likely to contain relevant information. For instance, for Java stack traces this leads to three groups: 1) top lines, which include the exception name and where the exception originated; 2) middle lines, which occur after the Java standard library traces and are most likely last lines of the application code closest to the bug; and 3) bottom lines, which can be useful for exceptions thrown from threads. Sampling from these three groups creates a generic recipe that  shortens the stack trace, captures different software designs, and preserves important information. Hence, for each stack trace, we decided to keep top 1 line, first 3 lines that refer to the application code, and bottom 1 line. Heuristic approaches such as this one have been reported to perform reasonably well even on unstructured runtime data (e.g., raw crash logs with multiple stack traces, possibly from different programming languages)~\cite{pradel2020scaffle}. %, hence we expect our heuristic to follow that trend given that we work only with Java stack traces.

For preprocessing code snippets, we decided to filter out punctuation for two reasons. First, in a recently published study, Paltenghi et al.~\cite{paltenghi2021thinking} compared the reasoning of developers and neural models, and observed that the models pay more attention to syntactic tokens (e.g., dots, periods, brackets), while developers focus more on strings or keywords. Given that developers perform better, DL models should mimic developers and put less attention to syntactic tokens. The second reason for filtering punctuation is pragmatic -- reducing the number of tokens to prevent exceeding the input limit size of the DL models.
Following preprocessing, each bug report is represented as a collection of OB, EB, S2R, stack traces, and code snippets.

\subsection{Natural language DA operators}

This group of operators is applied to OB, EB, and S2R due to their primarily natural language content. We propose to use two types of operators: token-level and paragraph-level.
Inspired by a simple yet effective technique called {\em Easy Data Augmentation} (EDA)~\cite{wei2019eda}, we propose to use 4 token-level operators.

\begin{itemize}
    \item \textbf{Dictionary Replace} - randomly selects a word from a pre-defined in-domain dictionary and replaces the word with its substitute.
    \item \textbf{Dictionary Insert} - works similarly to Dictionary Replace, however instead of replacing the word, this operator inserts the substitute at a random position in the text.
    \item \textbf{Random Swap} - randomly selects two words and swaps them.
    \item \textbf{Random Delete} - removes a randomly selected word.
\end{itemize}

\noindent
To build the in-domain dictionary for augmenting OB, EB and S2R, we use keywords from language patterns devised by Chaparro et al.~\cite{chaparro2017detecting}. The patterns specify combinations of different parts of speech with certain keywords that have to occur to classify a sentence or a paragraph as OB, EB or S2R. For instance, one of the most popular OB patterns is NEG\_VERB defined as: {\em (subject/noun phrase) ([adjective/adverb]) [negative verb] ([complement])}, where the negative verbs are defined as: {\em affect, break, block, close, etc.} 
The in-domain dictionary contains all keywords identified by Chapparo et al. and maps each keyword to its substitutes, e.g., {\em affect $\rightarrow$ \{break, block, close, ...\}}.
Domain knowledge guided operators have been recently shown to lead to better performance compared to more advanced but general approaches (e.g., embeddings)~\cite{kovatchev2021vectors}. 

%During augmentation, only keywords from the dictionary can be selected to be replaced or inserted a word. For instance, when augmenting OB that contains a negative verb, such as "disappear", Dictionary Replace can use a different negative verb from the predefined dictionary, e.g., "lack of". 

% https://drive.google.com/file/d/1KZRBWsPHS-XotuSYKuEm8_6eL4Hix3Sv/view?usp=sharing
\begin{figure*}
    \centering
    \includegraphics[width=\textwidth]{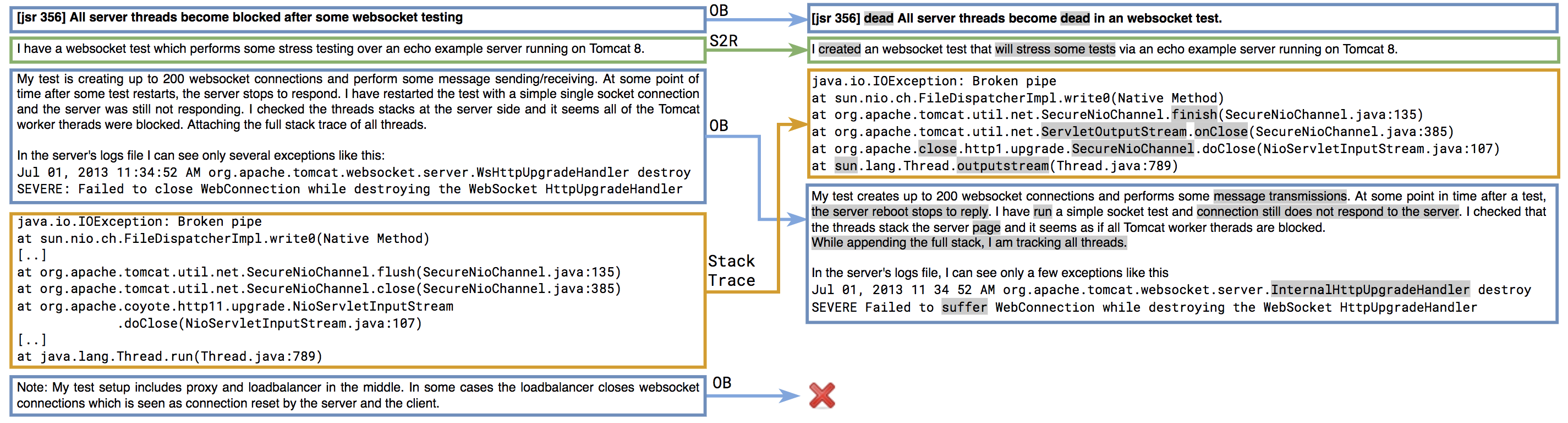}
    \caption{An example of augmented bug report for \texttt{Tomcat} \#55171. Token-level modifications are marked with grey color.}
    \label{fig:da-example}
\end{figure*}

As a paragraph-level operator, we use \textbf{Backtranslation} to translate paragraphs of OB, EB or S2R from English to German and back to English~\cite{ng2019facebook}. Backtranslation is a popular data augmentation operation that allows to paraphrase the original text.

Finally, let us describe how those operators are applied together to generate augmented data. For each bug report and for each OB, EB, S2R, we apply all token level operators $n$ times, where $n = \lambda * \#tokens$. \newtext{The operators are applied in the following order: replace, insert, swap and delete.} The value of $\lambda$ is set to $0.1$ for insert, replace, and swap operations, and $0.05$ for delete operation as these parameters have been empirically shown to produce best results~\cite{zou2020how}. Next, the Backtranslation operator is applied to paraphrase the modified text.
Given the randomness of the augmentation operators, the quality of the augmented sample may vary. \newtext{While, in general, natural language text typically retains its semantics when minor noise is introduced, bug reports are more structured type of data with certain keywords (e.g., code names), whose removal may have severe consequences for mapping the bug report to the source code.}
Hence, as a final step, we employ quality control that consists of two steps. First, we check if OB, EB and/or S2R can be still identified in the augmented paragraph using the BEE tool. For instance, if the original paragraph contained OB and EB, then the augmented version must contain OB and EB be to considered a valid paragraph. We also disallow changing the pattern (e.g., from OB to EB). Second, we ensure that no code tokens are lost during the augmentation by comparing the number of code tokens between the augmented and the original paragraph.

\subsection{Code-related DA operators}

In the context of this paper, code-related data refers to stack traces, code snippets, and code tokens present in natural language text. To augment code-related data, we propose 3 code token operators that are more strict versions of the natural language operators to minimize the risk of distorting the context.
\begin{itemize}
    \item \textbf{Code Token Replace} - randomly selects a code token and replaces it with its substitute.
    \item \textbf{Code Token Insert} - randomly selects a code token, and insert a substitute of that code token at a random position that is at most 3 positions away from the selected code token.
    \item \textbf{Code Token Swap} - swaps two randomly selected code tokens, such that (1) for stack traces code tokens can be swapped only between consecutive stack lines; (2) for code snippets, a swap operation must be performed within the surrounding 3 tokens \newtext{to minimize the potential distortion of the bug report's semantics. Changing this to a larger limit would result in more diverse samples being created, however, with a higher chance of disturbing the bug report's semantics.}. 
\end{itemize}
We decided against including a code token deletion operator as removing code tokens is more likely to disturb the invariance of augmented samples. 

\noindent
To find substitutes for a code token, first, for each bug report we build a dictionary of code names using class and method names that occur in its corresponding bug-inducing changesets. Next, we use the Levenshtein distance to measure the distance between the selected code token and all other tokens in the dictionary. \newtext{Levenshtein distance is a string similarity metric that quantifies similarity through the number of edits required to convert one string into another for strings of arbitrary length. We empirically observed that the most similar code tokens are often of the same form, and hence, may introduce very limited diversity into the augmented samples. For instance, consider the code token {\em word} with the top-3 closest tokens {\em is\_word}, {\em set\_word}, {\em get\_word}, and the top-20 token {\em check\_word\_missing\_letter}. To allow for more diverse augmentation, based on empirical observations for one of the evaluation projects, we opted for a less conservative top-$k$ selection, with $k$ set to 20}. Hence, a code token substitute is selected randomly from the 20 code tokens that have the lowest Levenshtein distance from the given code token. 
% \newtext{Note that one may want to choose a different value for $k$, where small value of $k$ lead to more precise and less diverse matches, and vice-versa for larger values of $k$.}

\subsection{Building augmented bug reports}
After augmentation, each bug report is decomposed into a collection of the original and augmented samples, i.e., natural language data (OB, EB, S2R), and code-related data (stack traces and code snippets). 
The remaining question is how to build a synthetic bug report out of all the available samples. 
%One possibility is to use each sample as a new synthetic bug report. However, this approach is imperfect since (1) a single sample (e.g., only OB) is likely to be less informative than a full bug report; and (2) a bug report that contains only OB, EB, S2R, stack trace \textbf{or} code snippets, occurs rarely in reality, thus creating a gap between training and testing data.
Recent work in neural machine translation has shown that concatenating augmented samples introduces structural diversity that prevents a DL model from learning to focus only on one part of the input, thus leading to a strong improvement in the model's performance~\cite{wu2021mixseq,nguyen2021data}.
We propose to use a similar approach to build augmented bug reports. More specifically, first we recreate the original structure of a bug report by concatenating augmented samples. Next, samples are reordered and at most 1 sample can be dropped to achieve further structural diversity. While dropping parts of bug reports may seem counterintuitive, DA strategies that remove tokens or sentences has been observed to have a positive impact on large pre-trained DL models ~\cite{chen2021hiddencut,shen2020simple}.
Figure~\ref{fig:da-example} shows bug report \#55171 from the \texttt{Tomcat} project and its augmented version. Each part of the bug report has been augmented separately using all of its respective DA operators. 
\newtext{In the case of natural language, we see semantically correct insertions and replacements (e.g., {\em blocked} $\rightarrow$ {\em dead}), while the paragraph rephrasing performed by Backtranslation is less precise yet still conveys the main message (e.g., the second OB in the figure). Code augmentation for this bug report includes augmenting stack trace and code tokens with code names from the bug inducing changeset. While such an approach inherently limits the possibility to add noise, it also introduces information about code components that are related to the bug report. This, in turn allows the model to learn these relations and utilize them during inference.}
Finally, When constructing the augmented bug report, the second OB and the stack trace have been swapped, while the third OB has been dropped, creating the final augmented version of bug report \#55171.

%This approach has the advantage of creating realistically looking bug reports with not only modified content but also diverse structure, hence better reflecting various styles of bug reports~\cite{?}.

\subsection{Ensuring a balanced augmented dataset}

% Bug localization data consists of pairs of bug reports and bug-inducing changesets, and each bug report is linked to at least one bug-inducing changeset. To create a training set from data that includes one-to-many relations, the data is flatten, such that one bug report creates one training sample with {\em each} of its bug inducing changesets.
% Moreover, given that a full changeset is typically a long document, and many of the popular DL models have a pre-defined input size limit (due to architectural decisions and/or performance), changesets are typically divided into smaller documents. Recent studies proposed using hunks, a set of consecutive line modifications that capture changes in one area of the file~\cite{ciborowska2021fast, wen2016}, as a implicit changeset division scheme. Hence, in training set a bug report that has $n$ introducing hunks creates $n$ training samples that repeats the content of that bug report, every time with a different introducing hunk. 

To increase the size of the bug localization training set with data augmentation, our approach is to focus on augmenting bug reports, increasing the number of pairs of bug reports and bug-inducing hunks. Recent studies show that using hunks, a set of consecutive line modifications that capture changes in one area of the file, produces improved retrieval results compared to using entire changesets~\cite{ciborowska2021fast, wen2016}. Hence, in this work, we build bug localization training set using pairs of bug reports and hunks extracted from bug-inducing changesets.
%Hence, in this work, we focus on localizing hunks instead of changeset.

Bugs affect different parts of source code base with varying frequency~\cite{catolino2019not}. In other words, parts of the source code (i.e., specific files or classes) are related to multiple bug reports and therefore their hunks can also be overrepresented in the original dataset. 
For instance, given a bug report with $n$ introducing hunks, data augmentation by a factor of 10 creates 10 new bug reports for each hunk, which leads to $10n$ new training samples. However, there is one major drawback to this DA approach. 
% This property is also reflected in bug localization datasets where one bug report can have multiple bug-inducing hunks, therefore occur in the training set multiple times. 
This data imbalance, created by the uneven distribution of bug reports and hunks in the training set, can be exasperated by DA, with strong downstream effects on the DL model and its prediction.

\begin{figure}
     \centering
     \begin{subfigure}[b]{0.9\columnwidth}
         \centering
         \includegraphics[width=0.95\columnwidth]{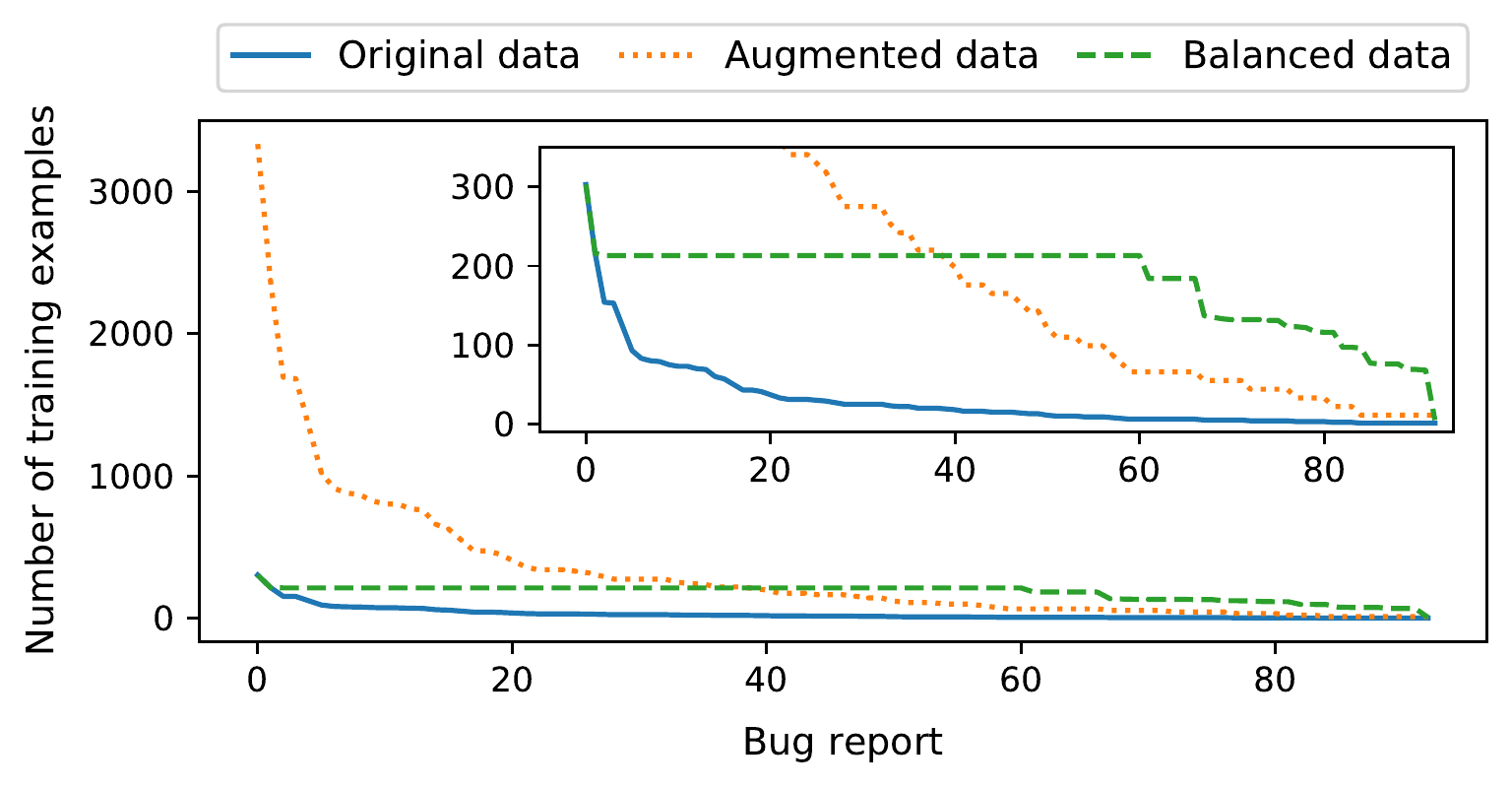}
         \caption{Number of times a bug report occurs in the training set.}
         \label{fig:imabalance-bugs}
     \end{subfigure}
     \begin{subfigure}[b]{0.95\columnwidth}
         \centering
         \includegraphics[width=0.89\columnwidth]{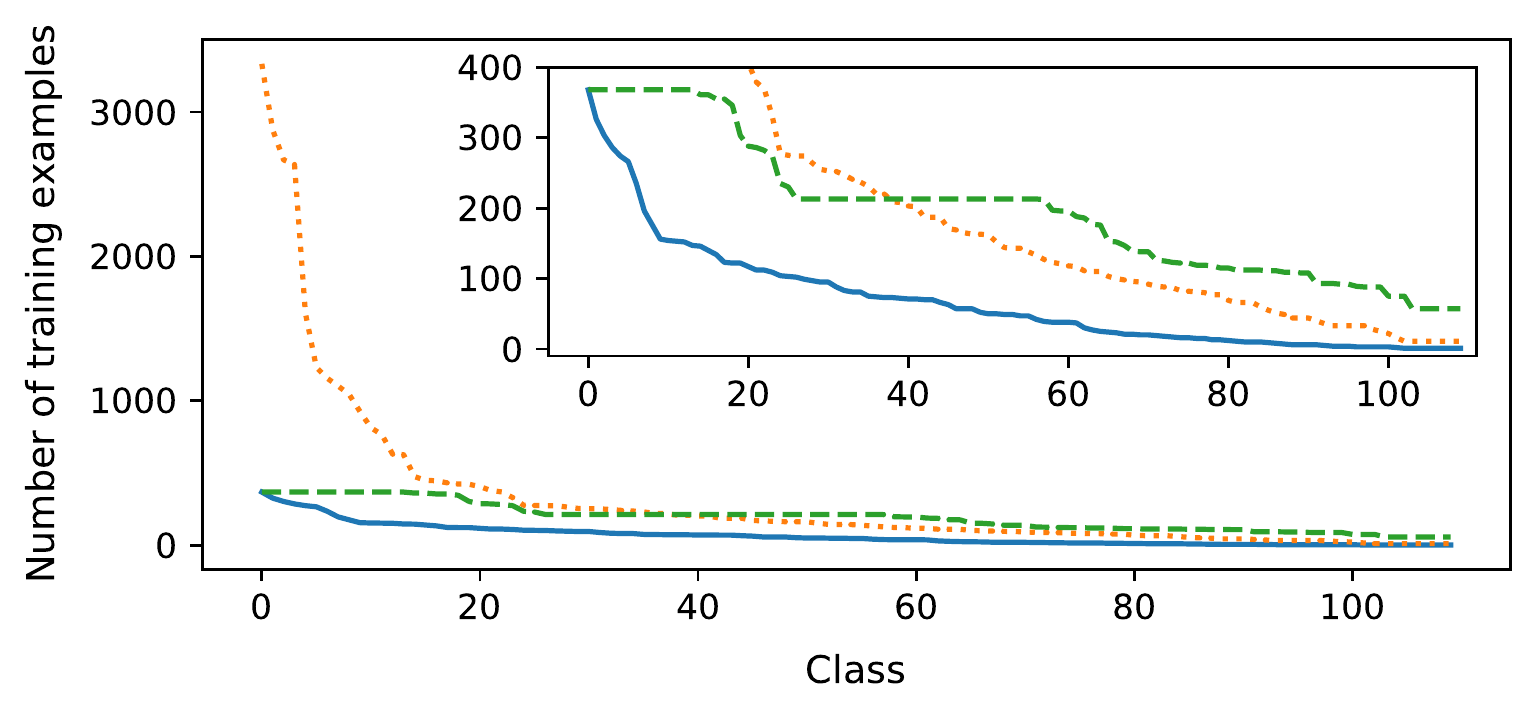}
         \caption{Number of times a class occurs in the training set.}
         \label{fig:imbalance-classes}
     \end{subfigure}
    \caption{Distribution of bug reports and classes showing data imbalance in bug localization training set.}
    \label{fig:imbalance}
\end{figure}

To provide further evidence, we empirically checked the dataset published by Wen et al.~\cite{wen2016}, which we use in our study.
Figure~\ref{fig:imbalance} shows the distribution of bug reports (Fig.~\ref{fig:imabalance-bugs}) and class occurrences (Fig.~\ref{fig:imbalance-classes}) for the \texttt{Tomcat} project. \newtext{The distributions shows how many samples in the training dataset refer to a specific bug report (or a class), where a bug report (or a class) has as many occurrences as the number of hunks.} 
We show these distributions for three different choices of training datasets: the original unaugmented dataset, a 10x augmented dataset, and an artificially balanced dataset. Within the plots, there are zoomed-in versions to increase readability at the smaller scale. In the plot for the original training set (i.e., the blue line), we observe that 11 out of 97 bug reports cover over 50\% (1432 out of 2812) of the training samples. \newtext{In other words, 50\% of the samples in the training set refer to 11 bug reports, since these bug reports have multiple introducing hunks, which translates to multiple entries in the training set. At the same time, 39 bug reports occur less than 10 times.} Similarly, out of the unique 110 classes that introduced a bug, the top 10 classes with most frequently occurring hunks cover 34.5\% (2586 out of 7478) of all training data. This imbalance in the training data can have two potential consequences for supervised training of a DL model. First, the model is more likely to learn the structure and semantics of bug reports that have a large number of bug inducing hunks, while neglecting less frequent bug reports. Secondly, classes that occur the most in the training set are more likely to be selected as bug-inducing by the trained model since they were often seen during training as bug-inducing. The issue of data imbalance has also been recognized in defect prediction datasets~\cite{yedida2021value}.

% While augmentation of all positive samples by a factor increases the size of the training set, it also amplifies the characteristics of the original data. In our case of handling imbalanced data, 
How augmentation exacerbates the problem of uneven data distribution can be observed in Figure~\ref{fig:imbalance}, where the orange dotted line depicts the data distributions in a dataset that was augmented by a factor of 10. The majority bug reports and classes become even more dominant in the augmented dataset, making the data imbalance problem more severe than in the original dataset.
To mitigate this problem, we propose a data balancing strategy that deliberately chooses samples to augment in order to smooth out the distributions of bug reports with respect to the source code. There are two main concerns that a data balancing strategy has to consider: (1) increasing the number of training samples for infrequent bug reports, and (2) ensuring that the number of samples with a given class does not dominate the dataset. To illustrate the need for these strategies, consider a bug report $B_1$ with 20 hunks from different classes, and a bug report $B_2$ with one hunk from class $C$. If the balancing strategy is focused only on the distribution of bug reports, then it creates 20 augmented samples for $B_2$, every time using the hunk from class $C$, hence $C$ is likely to be overrepresented in the training set. 
To address this, we introduce two augmentation factors $\alpha$ and $\omega$. While $\alpha$ influences the number of times each bug report is augmented, $\omega$ restricts how many times each class can be repeated in the augmented dataset.

%the proposed data balancing strategy takes into account the distribution of bug reports and classes when making a decision about adding a new augmented sample to the balanced dataset. 

% These two factors ($\alpha$ and $\omega$) are scaled by the maximum number of times a bug report occurs in the training dataset (i.e., the bug report that has the maximum number of hunks) and the maximum number a class is repeated in the training dataset. 

\begin{algorithm}
    \footnotesize
    \SetKwFunction{fun}{fun}
    \SetKwInOut{KwIn}{Input}
    \SetKwInOut{KwOut}{Output}

    \KwIn{$D_{train}$ -- training dataset;\\
          $\alpha$ -- augmentation factor;\\
          $\omega$ -- balancing factor}
    
    \KwOut{$D_{bl}$ -- balanced training dataset}

    $max_{br} \leftarrow \alpha  \times$ max. \# of bug reports in $D_{train}$ 
    
    $max_{cl} \leftarrow \omega  \times$ max. \# of classes in $D_{train}$ 

    $D_{bl} \leftarrow D_{train}$

    \For{$br$, $H_{br}$ in $D_{train}$}{
        \While{(count of $br$ in $D_{bal}) < max_{br}$}{  
            $br_a \leftarrow$ augment $br$ 
            
            $h_a \leftarrow$ select hunk $h_a$ from $H_{br}$, where
            
            %\hskip4.0em Min \#samples in $D_{bal}$ 
            
            \hskip4.0em (count of class for $h_a$ in $D_{bal}$)  $< max_{cl}$
            
            Add $br_{a}$, $h_{a}$ to $D_{bl}$
            }
    }

    \KwRet{$D_{bl}$}
    \caption{Data balancing with augmented bug reports}
    \label{alg:imb}
\end{algorithm}

The sequence of steps for the proposed data balancing augmentation strategy is presented in Algorithm~\ref{alg:imb}. 
In lines 1-2, we compute a limit for bug reports $max_{br}$ and classes $max_{cl}$ based on factors $\alpha$ and $\omega$ and the maximum number of times a unique bug report and class is present in the original training dataset.
Line 3 copies the existing data instances into the balanced dataset $D_{bl}$.
For each bug report that occurs below the $max_{br}$ limit, the algorithm augments the bug report (line 6), and selects a bug-inducing hunk from a class that occurs less than $max_{cl}$ times in $D_{bl}$ (lines 7-8), creating a new training sample. The algorithm continues to add new samples for a bug report until (1) the $max_{br}$ limit is reached, or (2) bug-inducing hunks from all the classes have reached $max_{cl}$. 
The result of this balancing strategy is depicted in the green line in Figure~\ref{fig:imbalance}, using values of $\alpha=0.7$ and $\omega=1.0$. Compared to the augmented dataset, the data distribution of the balanced dataset is obviously smoother, with a much more even representation of the source code.

\section{Experimental evaluation}

%In this section we describe the organization of our evaluation approach, while the results of the evaluation are presented in the next section. 

\subsection{Dataset and metrics}
To evaluate, we require a dataset that contains bug reports and bug-inducing changesets.
We use a dataset published by Wen et al.~\cite{wen2016} that contains manually-validated data from 6 open source software projects: AspectJ, JDT, PDE, SWT, Tomcat, and ZXing.  
Given that {\em infozilla} requires new lines to extract code snippets and stack traces, and new lines were removed from all bug reports in Wen et al.'s dataset, we located and re-scraped the bug reports (with new lines) from Bugzilla for all projects. For ZXing, the bug reports in the GitHub issue tracker did not match those collected by Wen et al., likely because the project was moved, and therefore ZXing was excluded from the evaluation set. 
To create a training set for each project, we ordered the bug reports by opening dates and selected the first half for training, while the remaining bug reports constitute the test set. 
%
% We used this unconventional 50-50 train-test split because we wanted to ensure a larger testing dataset.
%
Each positive training sample corresponds to a pair of a bug report and one of its inducing hunks (extracted from the inducing changeset). Each bug report includes the bug summary and description, while each hunk contains a log message and source code changes. 
The dataset of Wen et al. was constructed using SZZ~\cite{sliwerski2005when}, which identifies a changeset as bug-inducing if it shares {\em any} file modifications with bug fixing changeset. While a bug-inducing changeset may include modifications of multiple files, only a few of those may be relevant to a bug (as indicated by bug fixing changeset).
Hence, to ensure the quality of the training samples, we only include bug inducing hunks that refer to classes that also occurs in the bug fixing commit. 
For each positive sample, we create a negative sample by randomly selecting a hunk from a class which does not belong to the inducing changeset. After completing this step, for each project \newtext{we obtain our baseline dataset, $D_{ori}$.} 
%To answer the RQs, we created additional training datasets, which are described in details in the following sections.
The descriptive statistics of training and testing datasets used in this study are shown in Table~\ref{tab:dataset}. 
Note that the last column, {\em \# hunks}, denotes the number of \textit{all} hunks that are examined by the model during retrieval.

\begin{table}[t]
\small
\centering
\caption{Evaluation datasets.}
\label{tab:dataset}
\begin{tabular}{l|rr|rr}
\toprule
 & \multicolumn{2}{c|}{\textbf{Training}} & \multicolumn{2}{c}{\textbf{Testing}} \\ \midrule
 Project & \multicolumn{1}{c}{\# bugs} & \multicolumn{1}{c|}{\ori} &  \multicolumn{1}{c}{\# bugs} & \multicolumn{1}{c}{\# hunks} \\ \midrule
\textbf{AspectJ} & 100 & 2212  & 100 & 23446 \\
\textbf{SWT} & 45 & 9982 &  45 & 69833 \\
\textbf{Tomcat} & 96 & 5624  & 97 & 72134 \\
\textbf{PDE} & 30 & 3856 & 30 & 100373 \\
\textbf{JDT} & 47 & 18230 & 47 & 150630 \\
\bottomrule
\end{tabular}
\end{table}

% \begin{table}[]
% \scriptsize
% \centering
% \caption{Evaluation datasets}
% \label{tab:dataset}
% \begin{tabular}{p{0.8cm}|BDAAA|BH}
% \toprule
%  & \multicolumn{5}{c|}{\textbf{Training}} & \multicolumn{2}{c}{\textbf{Testing}} \\ \midrule
%  Project & \multicolumn{1}{B}{\# bugs} & \multicolumn{1}{D}{$D_{ori}$} & \multicolumn{1}{A}{$D_{neg}$} & \multicolumn{1}{A}{$D_{aug}$} & \multicolumn{1}{A|}{$D_{bl}$} & \multicolumn{1}{B}{\# bugs} & \multicolumn{1}{H}{\# hunks} \\ \midrule
% \textbf{AspectJ} & 100 & 2212 & 22120 & 24333 & 22420 & 100 & 23446 \\
% \textbf{JDT} & 47 & 18230 & 182300 & 200531 & 112944 & 47 & 150630 \\
% \textbf{PDE} & 30 & 3856 & 38560 & 42417 & 25058 & 30 & 100373 \\
% \textbf{SWT} & 45 & 9982 & 99820 & 109803 & 66886 & 45 & 69833 \\
% \textbf{Tomcat} & 97 & 5624 & 56240 & 61865 & 33816 & 97 & 72134 \\
% \bottomrule
% \end{tabular}
% \end{table}
To evaluate the retrieval performance of the DL models trained on different datasets, we use the following metrics.

\noindent
\textbf{Mean Reciprocal Rank:} MRR measures the retrieval accuracy using the reciprocal ranks of first relevant changeset in the ranking averaged across all bug reports. The higher the value of MRR is, the closer the bug-inducing changeset is to the top of the ranking.
$$MRR = \frac{1}{|B|} \sum_{i=1}^{|B|} \frac{1}{1stRank_{B_i}}.$$

\noindent
\textbf{Mean Average Precision:} MAP quantifies the ability of a model to retrieve all relevant changesets for a given bug report. 
MAP is calculated as the mean of Average Precision scores across all bug reports, where an Average Precision for a bug report is based on ranks of all relevant changesets in the ranking. 
The higher the values of MAP, the more relevant changesets is located in the top of the ranking.
$$MAP = \frac{1}{|B|} \sum_{i=1}^{|B|} \frac{1}{AvgP_{B_i}}.$$

\noindent
\textbf{Precision@K:} P@K measures how many of the top-$K$ changesets in the ranking are relevant to a bug report. The higher the value of P@K, the more relevant changesets can be found in top-$K$ positions.
%Similarly, as for MAP and MRR, the more the merrier.
$$P@n = \frac{1}{|B|} \sum_{i=1}^{|B|} \frac{|Rel_{B_i}|}{K}.$$

\subsection{DL models}
To evaluate the impact of the proposed data augmentation and balancing strategies on the retrieval performance, we train and evaluate three BERT-based~\cite{devlin2019bert} code retrieval architectures.

\noindent
\textbf{TBERT-Single}~\cite{lin2021traceability, dai2019deeper,nogueira2020passage} is the most straightforward approach for information retrieval with BERT. The model concatenates a bug report and a hunk, and processes it through BERT and a pooling layer to obtain a fused vector representation, which is subsequently passed to the classification head to obtain a relevancy score. While this model typically provides high retrieval accuracy, it also incurs significant retrieval delay, since a bug report needs to be compared with {\em all} hunks available in a project.

\noindent
\textbf{TBERT-Siamese}~\cite{lin2021traceability, reimers2019sentencebert} processes a bug report and a hunk sequentially through BERT and a pooling layer, creating two features vectors, that are subsequently concatenated and passed to the classification layer to produce the relevancy score. The key difference between TBERT-Single and TBERT-Siamese is in the opportunity to perform offline encoding of feature vectors for hunks, hence reducing the retrieval delay.

\noindent
\textbf{FBL-BERT}~\cite{ciborowska2021fast,khattab2020colbert} is a recently proposed BERT-based architecture that enables rapid retrieval across large collection of documents (i.e., hunks). Unlike TBERTs, which flattens the embedding matrix to a vector to make a prediction, FBL-BERT leverages the full embedding matrix and calculates relevancy score between a bug report and a hunk as a sum of maximum vector similarities between word embeddings of the bug report and hunk. This, in turn, allows to use efficient vector similarity search algorithms to find the most similar hunks and only re-rank those with FBL-BERT, hence significantly reducing the retrieval time per bug report. Given that FBL-BERT leverages fine-grained token-to-token embeddings matching, the model is more likely to better utilize relevant keywords if they occur in the bug report.

\noindent
\newtext{While all the models are based on BERT, their architectures differ in a few aspects. TBERT-Single concatenates a bug report and a changeset, processing them jointly through BERT, followed by a classification layer. TBERT-Siamese and FBL-BERT first use BERT to encode a bug report and a changeset separately, which results in two embedding matrices. The main difference between TBERT-Siamese and FBL-BERT is how they handle these matrices. TBERT-Siamese aggregates each matrix into a vector using a pooling operation, and, next, compares the embedding vector of a bug report with a changeset embedding vector. On the other hand, FBL-BERT uses both matrices to compute the relevancy score taking into account the embedding of each word.}

\subsection{Evaluation setup}
We performed the experiments on a server with Dual 12-core 3.2GHz Intel Xeon and 1 NVIDIA Tesla V100 with 32GB RAM memory running CUDA v.11.4. The models are implemented with PyTorch v.1.7.1, HuggingFace library v.4.3.2, and Faiss v.1.6.5 with GPU support. 
We opted for using BERTOverflow~\cite{tabassum2020code} as our pre-trained base BERT model, since, similarly to our data, StackOverflow data is also a mixture of code and natural language.
All models are fine tuned for 4 epochs, using a batch size of 16 and Adam  \newtext{(abbreviated from adaptive moment estimation)} optimizer \newtext{\cite{adamopt}} with learning rate set to 3e-6~\cite{devlin2019bert}.
Based on the average number of tokens in bug reports and hunks in our dataset, we set the input size limit to 256 and 512 tokens for bug reports and hunks respectively. All input documents are padded or truncated with respect to their input size limit.

\section{Results}

\begin{table}[t]
\small
\centering
\caption{Dataset characteristics for RQ1.}
\label{tab:rq1-datasets}
\begin{tabular}{l|rrrrr}
\toprule
 & \multicolumn{1}{c}{\textbf{AspectJ}} & \multicolumn{1}{c}{\textbf{SWT}} & \multicolumn{1}{c}{\textbf{Tomcat}} & \multicolumn{1}{c}{\textbf{PDE}} & \multicolumn{1}{c}{\textbf{JDT}} \\ \midrule
 \multicolumn{6}{c}{\textit{Not augmented datasets}} \\ \midrule
\textbf{\ori} & 2.2k & 9.9k & 5.6k & 3.9k & 18.2k \\
\textbf{\neg} & 22.1k & 99.8k & 56.2k & 38.6k & 182.3k \\ \midrule
 \multicolumn{6}{c}{\textit{Augmented datasets}} \\ \midrule
\textbf{\aug} & 24.3k & 109.8k & 61.9k & 42.4k & 200.5k \\
\textbf{\bla} & 22.4k & 66.9k & 33.8k & 25.1k & 112.9k \\
\textbf{\blb} & 29.8k & 90.5k & 46.8k & 30.7k & 142.3k \\
\textbf{\blc} & 31.5k & 95.1k & 49.0k & 32.5k & 150.5k \\
\textbf{\bld} & 44.2k & 130.9k & 65.6k & 46.7k & 216.4k \\
\bottomrule
\end{tabular}
\end{table}

\begin{table*}[]
\centering
\small
\caption{\newtext{Bug localization performance across all evaluation projects and for different training datasets: \ori -- original dataset; \neg -- dataset with 10x repeated instances; \aug -- augmented dataset; $D_{bl_{1-4}}$ -- augmented and balanced datasets. 
Statistically significant ($p < 0.05$) differences are marked with $\uparrow$ and $\downarrow$, where $\uparrow$ indicates significant improvement in performance compared to \ori, and $\downarrow$ indicates significant decrease compared to \ori.}}
\label{tab:rq1-general}
\begin{tabular}{l|lllll|lllll|lllll}
\toprule
 & \textbf{MRR} & \textbf{MAP} & \textbf{P@1} & \textbf{P@3} & \textbf{P@5} & \textbf{MRR} & \textbf{MAP} & \textbf{P@1} & \textbf{P@3} & \textbf{P@5} & \textbf{MRR} & \textbf{MAP} & \textbf{P@1} & \textbf{P@3} & \textbf{P@5} \\ \midrule
 & \multicolumn{5}{c}{\textbf{FBL-BERT}} & \multicolumn{5}{c}{\textbf{TBERT-Siamese}} & \multicolumn{5}{c}{\textbf{TBERT-Single}} \\ \midrule \midrule
 % & \multicolumn{15}{c}{\textit{All (\# bug reports = 331)}} \\ \midrule
$D_{ori}$ & 0.264          & 0.109          & 0.163          & 0.153          & 0.145          & 0.180          & 0.062          & 0.144          & 0.076          & 0.069          & 0.273          & 0.120          & 0.162 & 0.145 & 0.149 \\
$D_{rep}$ & 0.307\textsuperscript{$\uparrow$} & 0.129\textsuperscript{$\uparrow$} & 0.213\textsuperscript{$\uparrow$} & 0.179          & 0.176          & 0.201          & 0.086          & 0.110          & 0.093          & 0.093          & 0.271          & 0.140\textsuperscript{$\uparrow$} & 0.152 & 0.136\textsuperscript{$\uparrow$} & 0.176 \\
$D_{aug}$ & 0.353\textsuperscript{$\uparrow$} & 0.146\textsuperscript{$\uparrow$} & 0.247\textsuperscript{$\uparrow$} & {\bf 0.202\textsuperscript{$\uparrow$}} & 0.197\textsuperscript{$\uparrow$} & 0.236          & 0.103          & 0.157          & 0.124          & 0.119\textsuperscript{$\uparrow$} & 0.333\textsuperscript{$\uparrow$} & 0.144\textsuperscript{$\uparrow$} & 0.217\textsuperscript{$\uparrow$} & 0.188\textsuperscript{$\uparrow$} & 0.194\textsuperscript{$\uparrow$} \\
$D_{bl*}$ & {\bf 0.367\textsuperscript{$\uparrow$}} & {\bf 0.147\textsuperscript{$\uparrow$}} & {\bf 0.267\textsuperscript{$\uparrow$}} & 0.198\textsuperscript{$\uparrow$} & {\bf 0.206\textsuperscript{$\uparrow$}} & {\bf 0.328\textsuperscript{$\uparrow$}} & {\bf 0.107\textsuperscript{$\uparrow$}} & {\bf 0.247\textsuperscript{$\uparrow$}} & {\bf 0.150\textsuperscript{$\uparrow$}} & {\bf 0.146\textsuperscript{$\uparrow$}} & {\bf 0.368\textsuperscript{$\uparrow$}} & {\bf 0.149\textsuperscript{$\uparrow$}} & {\bf 0.269\textsuperscript{$\uparrow$}} & {\bf 0.192\textsuperscript{$\uparrow$}} & {\bf 0.182\textsuperscript{$\uparrow$}} \\  
\bottomrule
\end{tabular}
\end{table*}

\subsection{RQ1: (a) Can Data Augmentation improve the retrieval performance of DL-based bug localization? (b) How does Data Augmentation impact the performance of different DL-based bug localization approaches?}

\textbf{Setup.}
To evaluate the impact of DA on DL-based models, we compare the retrieval accuracy when training on the original, unaugmented dataset, \ori, to training with augmented and balanced data. More specifically, for each project we construct the five augmented datasets shown in Table~\ref{tab:rq1-datasets}.  \aug is an augmented, but unbalanced, dataset that contains 10 additional samples for each pair of bug report and hunk, while $D_{bl_i}$, $i=1,2,3,4$, are balanced datasets with different choices for ($\alpha$, $\omega$) = \{$(0.7, 1.0)$; $(0.85, 2.0)$; $(1.0, 2.0)$; $(1.3, 2.0)$\} respectively. The rationale for these specific values of $\alpha$ and $\omega$ is to explore $\alpha$ (in ranges that do not generate more data than we can manage computationally), while selecting values for $\omega$ that do not constrain $\alpha$'s effect.  
Given that augmentation increases the number of positive samples, the number of negative samples grows proportionally as well (i.e., for each positive sample, we randomly create one negative sample).
To ensure that the difference in performance is in fact the result of DA, and 
not the diversity introduced by a new negative samples,
%not the higher number of data instances, 
we created an additional baseline, \neg, that {\em repeats} positive samples without augmentation 10 times, and, correspondingly, also adds 10 new negative samples. In effect, the only difference between \neg and \aug is the fact that \aug uses augmented bug reports while \neg repeats the positive samples.
% Given that augmentation increases the number of positive samples, the number of negative samples grows proportionally as well (i.e., for each positive sample, there is one negative sample).
% To ensure that the difference in performance is in fact the result of DA (i.e., the new positive samples), and not the higher number of negative data points, we created an additional baseline, \neg, that repeats positive samples without augmentation 10 times, and, hence, also adds 10 new negative samples. In effect, the only difference between \neg and \aug is the fact that \aug uses augmented bug reports while \neg repeats the positives samples.

\noindent
All models are evaluated on the same (unaugmented) test set. Since TBERT-Single requires considerably more time than the other models (e.g., TBERT-Single takes more than 24h to run on the JDT project), 
we only evaluate it on one of the balanced datasets - $D_{bl_1}$ - as it exhibits the best performance for TBERT-Siamese, which uses a relatively similar DL architecture to TBERT-Single.
%we only evaluate it on one of the balanced datasets - $D_{bl_1}$ - as it has the smallest number of instances; $D_{bl_1}$ also exhibits the best performance for TBERT-Siamese and given the similarity of the two models' architectures. 

\noindent
To evaluate the statistical significance of the difference in performance when training DL models with and without data augmentation, we use the \textit{Student’s paired t test} to compute $p$-values between performance metrics of \ori and all other datasets (i.e., \neg, \aug, \blx)~\cite{smucker2007comparison,urbano2019staistical}. The test assumes the performance values to be normally distributed. We consider $p < 0.05$ to be statistically significant.

\smallskip
\noindent
\textbf{Results.}
Table~\ref{tab:rq1-general} shows the retrieval performance of FBL-BERT, TBERT-Siamese and TBERT-Single trained on four dataset: \ori, \neg, \aug and \blx, where \blx denotes the average best performing balanced dataset for the given model. \newtext{The top part of the table shows results across bug reports from all projects, followed by per project results.} 
In general, we observe that the models improve across all the metrics compared to \ori, with the lowest improvement noted for \neg, followed by \aug, and with the highest improvement recorded for \blx.

Depending on the model the scale of the improvement varies. 
While the MRR score for FBL-BERT increases from 0.264 for \ori to 0.367 for \blx, about half of the improvement can be attributed to the dataset size as indicated by the results for \neg with the MRR score of 0.307.
Moreover, we also observe that \aug improves the score from 0.307 for \neg to 0.353, indicating that using an augmented dataset makes a difference not only through data quantity. %. but also by data quality. 
The improvement between \aug and \blx is marginal and equal to 0.014, indicating that even the best balancing configuration has a small effect on FBL-BERT in general.

Training with a balanced dataset has a bigger impact on TBERT-Single and TBERT-Siamese with an improvement of 0.035 and 0.092 in MRR scores respectively when compared to \aug. \newtext{Moreover, data balancing is the key contributor to the improvement in TBERT-Siamese for which statistically significant difference is observed only for \blx. In the case of TBERT-Single, training the model with both \aug and \blx leads to significant improvement across all the metrics.}

%Finally, comparing the results of \neg and \aug for both TBERT models and FBL-BERT, we observe that \aug increases the retrieval accuracy across all metrics, indicating that the proposed bug reports augmentation approach is effective.

%The fact that we observe improvement even with \neg is encouraging in a sense that building \neg dataset is simple and straightforward, hence can be used as an initial check to see if the model's performance is affected by training data scarcity, before looking into more advanced approaches.

\begin{tcolorbox}
{\em RQ1 (a): Data augmentation improves the DL-based bug localization results across all models. Using data balancing with augmentation can further improve performance.
}
\end{tcolorbox}

% \begin{table}[t]
% \centering
% \caption{Bug localization performance for different training datasets.}
% \label{tab:rq1-general}
% \begin{tabular}{l|rrrrr}
% \toprule
%  & \multicolumn{1}{c}{\textbf{MRR}} & \multicolumn{1}{c}{\textbf{MAP}} & \multicolumn{1}{c}{\textbf{P@1}} & \multicolumn{1}{c}{\textbf{P@3}} & \multicolumn{1}{c}{\textbf{P@5}} \\ \midrule
% \textbf{Dataset} & \multicolumn{5}{c}{\textit{FBL-BERT}} \\ \midrule
% \ori & 0.264 & 0.109 & 0.163 & 0.153 & 0.145 \\
% \neg  & 0.307 & 0.129 & 0.213 & 0.179 & 0.176 \\
% \aug  & 0.353 & 0.146 & 0.247 & {\bf 0.202} & 0.197 \\
% \blb & {\bf 0.367}  & {\bf 0.147} & {\bf 0.267} & 0.198 & {\bf 0.206} \\ \midrule
%  & \multicolumn{5}{c}{\textit{TBERT-Siamese}} \\ \midrule
% \ori & 0.180 & 0.062 & 0.144 & 0.076 & 0.069 \\
% \neg & 0.201 & 0.086 & 0.110 & 0.093 & 0.093 \\
% \aug  & 0.236 & 0.103 & 0.157 & 0.124 & 0.119 \\
% \bla & {\bf 0.328} & {\bf 0.107} & {\bf 0.247} & {\bf 0.150} & {\bf 0.146} \\ \midrule
%  & \multicolumn{5}{c}{\textit{TBERT-Single}} \\ \midrule
% \ori & 0.273 & 0.120 & 0.162 & 0.145 & 0.149 \\
% \neg & 0.271 & 0.140 & 0.152 & 0.136 & 0.176 \\
% \aug  & 0.333 & 0.144 & 0.217 & 0.188 & {\bf 0.194} \\
% \bla & {\bf 0.368} & {\bf 0.149} & {\bf 0.269} & {\bf 0.192} & 0.182 \\
% \bottomrule
% \end{tabular}
% \smallskip
% \hfill\parbox[t]{8cm}{\smallskip \scriptsize \ori -- original dataset; \neg -- dataset with 10x repeated instances;\\ \aug -- augmented dataset; $D_{bl_{1-4}}$ -- augmented and balanced datasets;}
% \end{table}

\begin{table}[t]
\small
\centering
\caption{Bug localization performance with different augmented and balanced training datasets.}
\label{tab:rq1-balanced}
\begin{tabular}{l|rrrrr|rrr}
\toprule
 & \multicolumn{1}{c}{\textbf{MRR}} & \multicolumn{1}{c}{\textbf{MAP}} & \multicolumn{1}{c}{\textbf{P@1}} & \multicolumn{1}{c}{\textbf{P@3}} & \multicolumn{1}{c|}{\textbf{P@5}} & $\alpha$ & $\omega$ & \#$D$\\ \midrule
\textbf{Dataset} & \multicolumn{7}{c}{\textit{FBL-BERT}} \\\midrule
\bla & 0.314 & 0.128 & 0.210 & 0.183 & 0.177 & 0.70  & 1.0 & 260k\\
\blb & {\bf 0.367} & 0.147 & {\bf 0.267} & 0.198 & 0.206 & 0.85 & 2.0 & 340k\\
\blc & 0.357 & {\bf 0.155} & 0.260 & {\bf 0.204} & {\bf 0.215} & 1.00  & 2.0 & 360k\\
\bld & 0.315 & 0.142 & 0.217 & 0.176 & 0.179 & 1.30 & 2.0 & 500k\\ \midrule
 & \multicolumn{7}{c}{\textit{TBERT-Siamese}} \\ \midrule
\bla & {\bf 0.328} & {\bf 0.107} & {\bf 0.247} & {\bf 0.150} & {\bf 0.146} & 0.70  & 1.0 & 260k\\
\blb & 0.220 & 0.080 & 0.140 & 0.116 & 0.111 & 0.85 & 2.0 & 340k\\
\blc & 0.215 & 0.081 & 0.130 & 0.105 & 0.117 & 1.00 & 2.0 & 360k\\
\bld & 0.182 & 0.068 & 0.107 & 0.091 & 0.086 & 1.30 & 2.0 & 500k\\
\bottomrule
\end{tabular}
\end{table}

Table~\ref{tab:rq1-balanced} shows the retrieval accuracy for FBL-BERT and TBERT-Siamese when the models are trained on different balanced datasets, with the values of $\alpha$, $\omega$ and the dataset size provided on the right side of the table. In the case of FBL-BERT, \blb and \blc provide on average the best performance, improving MRR and MAP by 16.8\% and 14.8\% compared to \bla and \bld. However, as noted before, the improvement over the imbalanced dataset \aug is marginal.  
In case of TBERT-Siamese, the smallest balanced dataset, \bla, produces the highest MRR score of 0.328 which outperforms other balanced dataset by at least 49\%.

% To better understand how different balanced datasets affect the models' performance, in Figure~\ref{fig:rq1-imbalance} we show MRR scores across all evaluation projects ordered by the project sizes, i.e., the number of hunks in a project (see Table~\ref{tab:dataset} for details). Interestingly, for FBL-BERT we observe that the larger the project, the more improvement when training with the bigger training dataset. More specifically, while for AspectJ, SWT, and Tomcat the maximum MRR scores are obtained with \blb, PDE performs best with \blc, and JDT with \bld.
% On the other hand, TBERT-Siamese consistently achieves the highest MRR scores for all projects with \bla, while other datasets, albeit larger, do not bring improvement. 

\begin{figure}[t]
     \centering
     \begin{subfigure}[b]{0,49\columnwidth}
         \centering
         \includegraphics[width=\columnwidth]{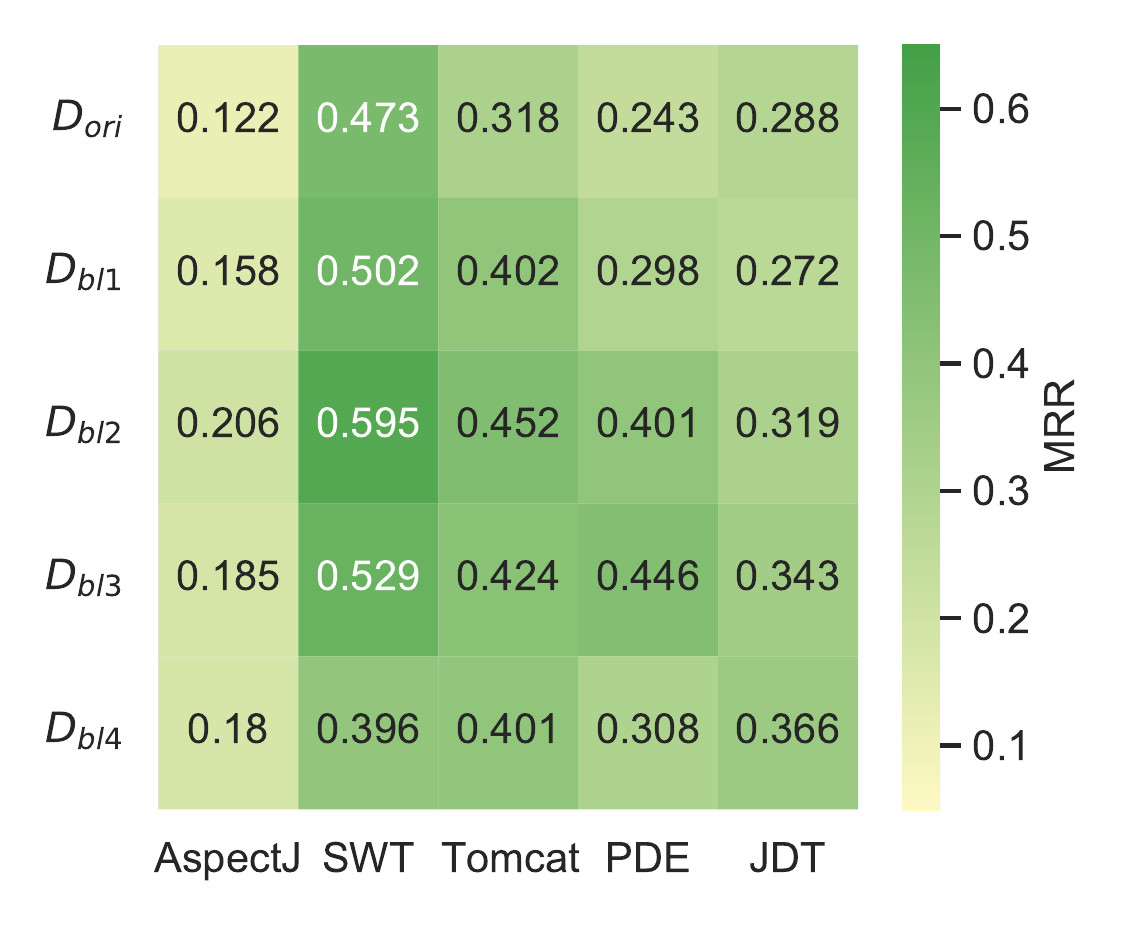}
         \caption{FBL-BERT}
         \label{fig:rq1-fbl}
     \end{subfigure}
     \begin{subfigure}[b]{0.49\columnwidth}
         \centering
         \includegraphics[width=\columnwidth]{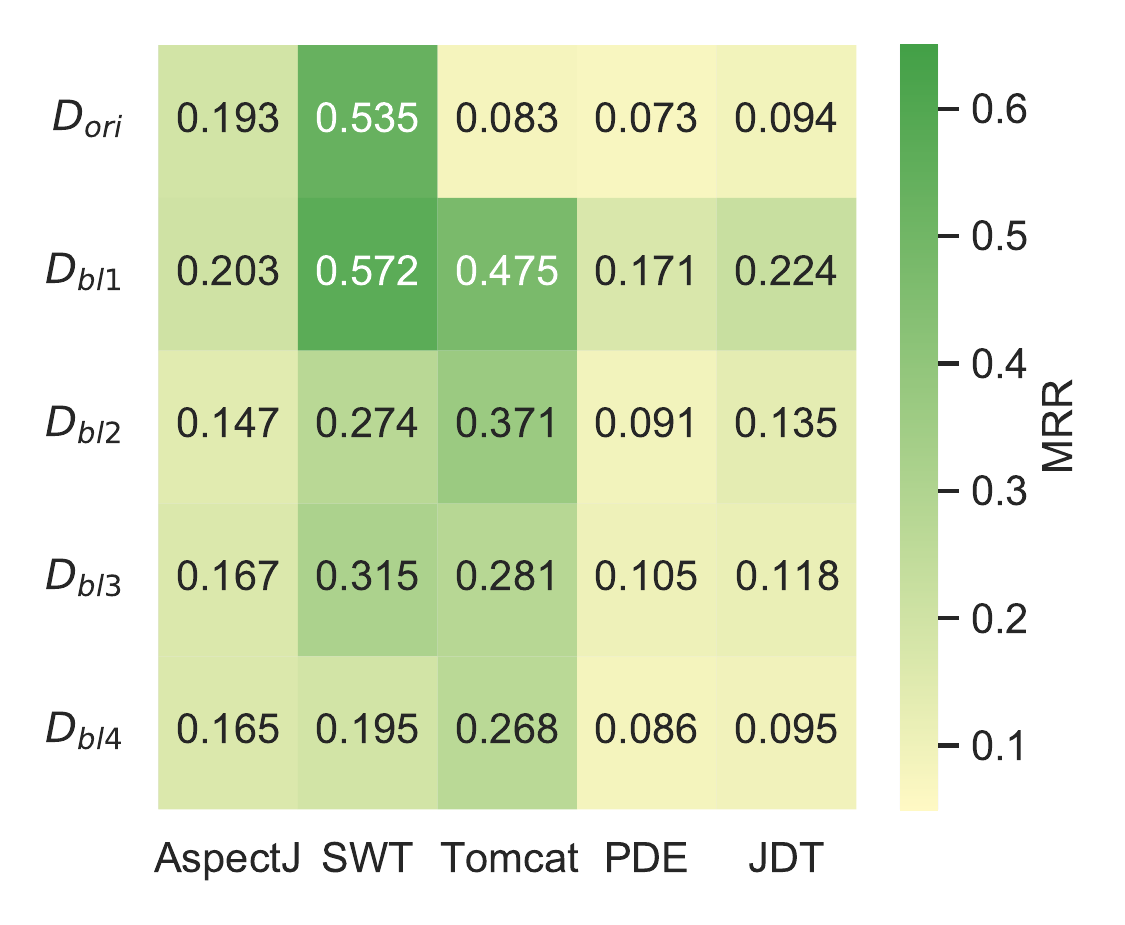}
         \caption{TBERT-Siamese}
         \label{fig:rq1-siamese}
     \end{subfigure}
    \caption{MRR scores for evaluation projects trained on different balanced datasets.}
    \label{fig:rq1-imbalance}
\end{figure}

The difference in the models' performance with different datasets can be attributed to (1) the overall difference in the models' architectures affecting models demand for training data; and (2) the size of each project measured as the number of hunks (see Table~\ref{tab:dataset}).
To better understand if and when model may require more data, in Figure~\ref{fig:rq1-imbalance} we show MRR scores across all evaluation projects ordered by their size, i.e., the number of hunks in a project.

\begin{tcolorbox}
{\em RQ1 (b): The results indicate that different model architectures may have different needs in terms of training dataset size to achieve their optimal performance. Some models benefit from more augmented samples, especially for larger projects.}
\end{tcolorbox}

\subsection{RQ2: Which of the proposed DA operators contribute the most to retrieval performance?}

\textbf{Setup.}
To better understand the influence of the proposed data augmentation operators on the downstream model effectiveness, we perform ablation studies on training datasets created using all but one augmentation operator type. To this end, we create 5 types of augmented training datasets: No Backtranslation, No Insert, No Delete, No Replace and No Swap operator. Note that we consider, e.g., both Code Token Swap and Random Swap as operators of Swap type. To balance the datasets, we use $\alpha$ and $\omega$ values from RQ1 that resulted in the best performance for the models, i.e., for FBL-BERT $\alpha=0.85$, $\omega=2.0$, while for TBERT models $\alpha=0.7$, and $\omega=1.0$.

\begin{figure}[t]
    \centering
    \includegraphics[width=\columnwidth]{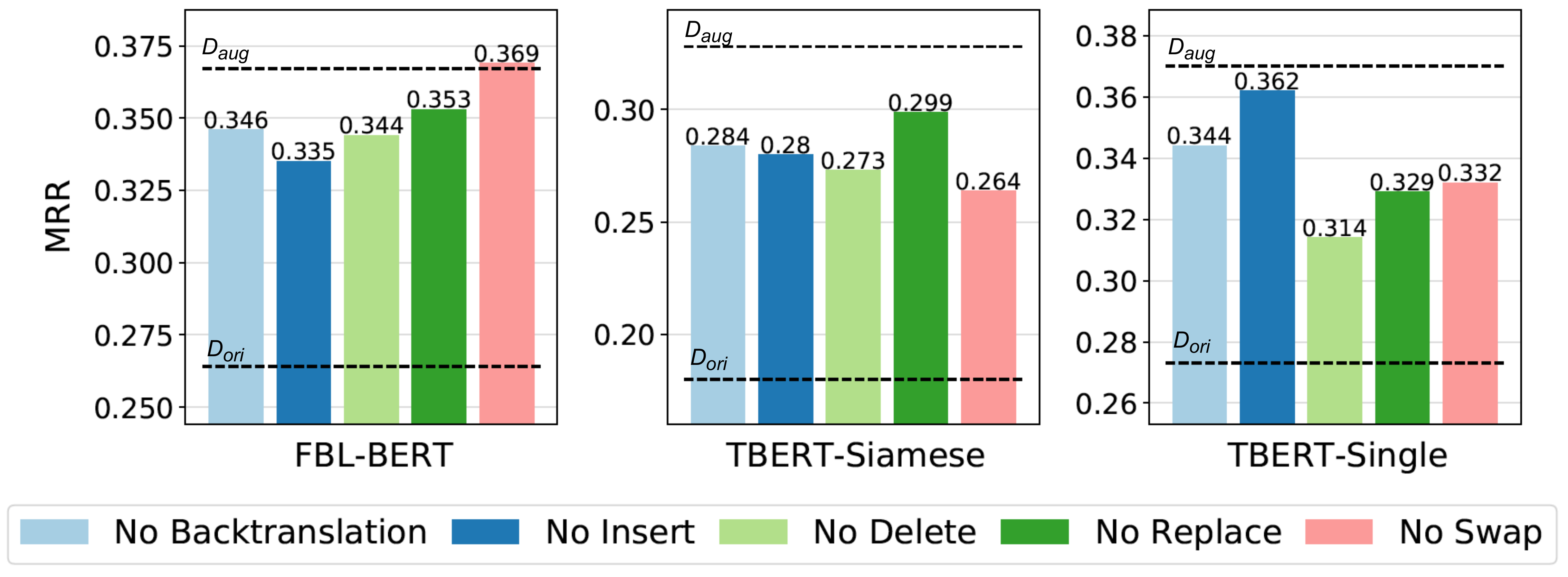}
    \caption{MRR scores when trained with augmented data using different DA operators.}
    \label{fig:rq2}
\end{figure}

\noindent
\textbf{Results.}
Figure~\ref{fig:rq2} shows the MRR scores for datasets augmented with 4 out of 5 operator types as well as MRR scores of \ori and \aug as horizontal lines for reference. We note that most of the operators contribute towards the final performance, with an exception of Swap operator for FBL-BERT. The lack of impact for Swap operator can be attributed to the model architecture. Given that FBL-BERT leverages all of the tokens in a bug report separately, swapping the token positions does not preclude them from being matched. On the other hand, excluding Random Insert affects FBL-BERT the most, indicating that inserted tokens are valuable to the model and improve its effectiveness when matching token embeddings.
The Delete operator is the most prominent contributor to the performance of both TBERT models. When the Delete operator is disallowed during augmentation, the MRR score of augmented datasets drops by 0.054 and 0.055 for TBERT-Single and TBERT-Siamese respectively, indicating that the variance caused by removing tokens randomly has a positive impact. \newtext{The Delete operator, when applied in relative moderation, seems to add to the robustness of the models, i.e., the models create additional links between terms and concepts in the bug reports and changesets. The value of this process is also supported by recent work in adversarial training of large language models, like BERT, in order to improve their robustness against malicious attacks~\cite{hsieh-etal-2019-robustness}.}%, especially when combined with the Quality Control module.

\begin{tcolorbox}
{\em RQ2: All DA operators contribute to the performance improvement with varying degree, with the exception of Swap for FBL-BERT. The Delete operator consistently improves performance in all three models.}
\end{tcolorbox}

% \begin{table}[t]
% \centering
% \caption{Retrieval performance when trained with augmented data using different operators.}
% \label{tab:rq2}
% \begin{tabular}{l|lllll}
% \toprule
%  & \textbf{MRR} & \textbf{MAP} & \textbf{P@1} & \textbf{P@3} & \textbf{P@5} \\  \midrule
% \textbf{Dataset} & \multicolumn{5}{c}{\textit{FBL-BERT}} \\  \midrule
% Original & 0.367 & 0.147 & 0.267 & 0.198 & 0.206 \\
% NBACK & 0.346 & 0.153 & 0.240 & 0.201 & 0.201 \\
% NINSERT & 0.335 & 0.146 & 0.210 & 0.212 & 0.215 \\
% NRD & 0.344 & 0.146 & 0.247 & 0.194 & 0.202 \\
% NREPLACE & 0.353 & 0.161 & 0.243 & 0.214 & 0.221 \\
% NRS & 0.369 & 0.159 & 0.257 & 0.209 & 0.224 \\  \midrule
%  & \multicolumn{5}{c}{\textit{TBERT-Siamese}} \\  \midrule
% Original & 0.328 & 0.107 & 0.247 & 0.150 & 0.146 \\
% NBACK & 0.284 & 0.090 & 0.194 & 0.133 & 0.129 \\
% NINSERT & 0.280 & 0.088 & 0.190 & 0.140 & 0.133 \\
% NRD & 0.273 & 0.094 & 0.180 & 0.142 & 0.133 \\
% NREPLACE & 0.299 & 0.104 & 0.217 & 0.141 & 0.141 \\
% NRS & 0.264 & 0.098 & 0.170 & 0.130 & 0.126 \\  \midrule
%  & \multicolumn{5}{c}{\textit{TBERT-Single}} \\  \midrule
% Original & 0.368 & 0.149 & 0.269 & 0.192 & 0.182 \\
% NBACK & 0.344 & 0.160 & 0.224 & 0.196 & 0.216 \\
% NINSERT & 0.362 & 0.161 & 0.259 & 0.210 & 0.207 \\
% NRD & 0.314 & 0.140 & 0.207 & 0.167 & 0.175 \\
% NREPLACE & 0.329 & 0.154 & 0.224 & 0.186 & 0.194 \\
% NRS & 0.332 & 0.142 & 0.217 & 0.174 & 0.191 \\
% \bottomrule
% \end{tabular}
% \end{table}

\subsection{Threats to validity}
There are several validity threads of our findings.
A threat to internal validity of the study are the parameter choices for DL-based bug localization models, particularly in the context of (1) training procedure; (2) BERT-base selection; and (3) parameters inherent to each model. 
To mitigate that threat, during training we follow recommendations of BERT authors~\cite{devlin2019bert}, while for each model we use parameters identified as optimal by the previous studies~\cite{lin2021traceability, ciborowska2021fast}. While in our study, we use BERTOverflow as our BERT base model, other choices exist (e.g., CodeBERT~\cite{feng2020codebert}), and more models are underway, hence we leave the evaluation of different BERT base models in the context of bug localization to future work.

%A threat to the internal validity of the study are the parameter choices for the DL-based bug localization models.
%A mitigating factor is our use of parameters considered to be optimal in prior uses of BERT-based models
%\cite{khattab2020colbert,devlin2019bert}. 

Another internal threat is in our choices of augmentation operators, their parameters (e.g., $\lambda$), and how they are applied together (e.g., stacking operators). This threat is mitigated by following best known practices from the NLP augmentation literature that focus on token-level operators and in-domain tasks~\cite{wei2019eda,kovatchev2021vectors}.
While we explore some parameter choices for data balancing in the paper (e.g., $\alpha$ and $\omega$), there are also additional parameters related to bug report building process as well as other possible augmentation operators that may provide more improvement. 

The external tools we leveraged to build our augmentation pipeline, e.g., infozilla, BEE tool, can introduce noise that propagates to our reported results. However, these are state-of-the-art tools that have been thoroughly evaluated so their error rate should be limited. 

Furthermore, the randomness of the augmentation operators may pose a threat to the internal validity. To mitigate that, we ensure to set an initial value on the system’s pseudo-random number generator when building an augmented dataset as well as when training a DL model.

A threat to the external validity is that we evaluated the data augmentation technique only for bug localization on a limited number of bugs collected from a selection of open source Java projects. This threat is mitigated by the fact that the dataset has been used in several prior bug localization studies ~\cite{saha_bluir_2013,wong_brtracer_2014,wen2016}. Another mitigating factor is that the projects reflect a variety of purposes, development styles and histories.

Limitations in the chosen evaluation metrics pose a threat to conclusion validity as they may not directly measure user satisfaction with the retrieved change hunks~\cite{wang2015usefulness}. The threat is mitigated by the fact that the selected metrics are well-known and widely accepted as best available to measure and compare the performance of IR techniques.

\section{Conclusion and Future Work}
DL models toward bug localization excel at bridging the lexical gap between natural language describing a bug report and programming language that defines the source code. However, training an effective DL model requires large amount of project-specific labelled data (i.e., pairs of bug reports and bug-inducing changesets), that is typically difficult to obtain in sufficient quantity for a single project.
To relax the requirement on data quantity, and enable using DL model when training data is scarce, this work proposes to use data augmentation (DA) to create new, realistically looking bug reports that can be used to significantly increase the size of the training set. To augment bug reports, we propose DA operators that independently augment the natural language and code-related content of a bug report. To build a new training dataset using augmented bug reports, we propose a data balancing strategy that selectively augments bug reports to add more training samples for underrepresented parts of the source code. 

The results indicate that the proposed data augmentation improves retrieval accuracy across all studied DL models increasing MRR score by 39\% to 82\% compared to the original, unaugmented dataset. Moreover, when augmented datasets are compared against training sets expanded by data repetition, we observe that they improve MRR scores by 20\% to 36\%.
All of the proposed DA operators contribute to the final performance, with token deletion bringing most consistent impact for different DL models.

\newtext{This is one of the first papers to introduce data augmentation for software engineering. We believe data augmentation as a technique has potential for SE because the datasets are not as large as in mainstream ML. In addition, data augmentation is not a one-size-fits-all technique and its optimal application requires custom operators, so the paper contributes in designing the data augmentation operators for bug reports and applying them using a data balancing strategy.}

Despite this, the proposed approach requires more experiments to strengthen our observations and recommendations. As our future work, we plan to (1) extend our evaluation datasets with new software projects written in Java, Python and Javascript; (2) conduct experiments with more heavily augmented data, i.e., by using DA operators on a larger number of tokens; (3) experiment with different deletion operators (e.g., removing irrelevant code tokens) given the good performance of Random Delete for natural language; and (4) experiment with different configurations for the bug report builder.

\section{Data Availability}
A replication package that includes all relevant code and scripts is available at \url{https://anonymous.4open.science/r/fbl-bert-987B}.

\bibliographystyle{ACM-Reference-Format}
\bibliography{main}

\end{document}